\documentclass[twocolumn]{aa}
\usepackage{graphicx}
\usepackage{txfonts}

\newcommand{\myrule}{\rule[-0.1cm]{0.cm}{0.7cm}} 

\begin{document}
   \title{Improved kinematics for brown dwarfs \\and very low-mass stars in Cha\,I\\
and a discussion of brown dwarf formation 
          \thanks{Based on observations at the Very Large Telescope of the 
European Southern Observatory at Paranal, Chile 
in program 65.L-0629, 65.I-0011, 268.D-5746, 72.C-0653.}
         }

   \titlerunning{Improved kinematics for brown dwarfs in Cha\,I
                }

   \subtitle{}

   \author{V. Joergens
          \inst{1}
          }

   \offprints{V. Joergens, \email{viki@strw.leidenuniv.nl}}

   \institute{Leiden Observatory / Sterrewacht Leiden,
              P.O.Box 9513, 2300 RA Leiden, Netherlands
              }

   \date{Received 12 April 2005; accepted 14 September 2005}

   \abstract{
We present a precise kinematic study of very young brown dwarfs in
the Cha\,I cloud based on radial velocities (RVs) measured with UVES at the VLT.
The kinematics of the brown dwarfs in Cha\,I are compared to the 
kinematics of T~Tauri stars in the same field, based on both
UVES measurements for very low-mass ones and on RVs from the literature.
More UVES spectra were taken compared with a former paper (Joergens \& Guenther 2001),
and the reduction of the spectra was improved, while   
studying the literature for RVs of T~Tauri stars in Cha\,I 
led to a cleaned and enlarged sample of T~Tauri stars. 
The result is an improved empirical RV distribution of brown dwarfs, as well as 
of T~Tauri stars in Cha\,I.
We found that the RVs of the nine brown dwarfs and very low-mass stars (M6--M8)
in ChaI that were studied have a mean value of 15.7\,km\,s$^{-1}$ and a 
dispersion measured in terms of a standard deviation of 0.9\,km\,s$^{-1}$, 
and they cover a total range of 2.6\,km\,s$^{-1}$.
The standard deviation is consistent with the dispersion
measured earlier in terms of fwhm of 2.1\,km\,s$^{-1}$.
The studied sample of 25 T~Tauri stars (G2--M5) has a mean RV of 14.7\,km\,s$^{-1}$,
a dispersion in terms of standard deviation of 1.3\,km\,s$^{-1}$ and
in terms of fwhm of 3.0\,km\,s$^{-1}$, and a total range of 4.5\,km\,s$^{-1}$.
The RV dispersion of the brown dwarfs is consistent within the errors with that 
of T~Tauri stars, which is in line with the finding of no mass dependence in some
theoretical models of the ejection-scenario for the formation of brown dwarfs.
In contrast to current N-body simulations, we did not find a high-velocity tail for the 
brown dwarfs RVs.
We found hints suggesting different kinematics for binaries compared to
predominantly single objects in Cha\,I, as suggested by some models.
The global RV dispersion for Cha\,I members 
(1.24\,km\,s$^{-1}$) is significantly lower than for Taurus members (2.0\,km\,s$^{-1}$), 
despite higher stellar density in Cha\,I showing that 
a fundamental increase in velocity dispersion with stellar density of the star-forming region 
is not established observationally.
The RVs of brown dwarfs observed in Cha\,I are less dispersed than predicted by existing models
for the ejection-scenario.

   }

   \maketitle
%

\section{Introduction}

The formation of objects below or close to the hydrogen burning 
limit is one of the main open issues in the field of the origins of solar systems.
The almost complete absence of brown dwarfs in close ($<$3\,AU) orbits around solarlike stars 
(`brown dwarf desert') found in ongoing high-precision RV surveys compared to the detection of more 
than 150 extrasolar planets in this separation range 
(e.g. Moutou et al. 2005; Marcy et al. 2005)
suggests that brown dwarfs generally do not form like planets by dust condensation in a circumstellar 
disk.
On the other hand, a starlike formation from direct gravitational collapse and fragmentation 
of molecular clouds requires the existence of cloud cores that are cold and dense enough to become 
Jeans-unstable for brown dwarf masses, which have not yet been found. 
On theoretical grounds, the opacity limit for the fragmentation (Low \& Lynden-Bell 1976)
might prevent the formation of (lower mass) brown dwarfs by direct collapse.

An alternative scenario was proposed in recent years, namely the formation of 
brown dwarfs by direct collapse of unstable cloud cores of stellar masses that would 
have become stars if the accretion was not stopped at an early 
stage by an external process before the object had accreted to stellar mass.
It was proposed that such an external process could be the ejection of the 
protostar out of the dense gaseous environment due to dynamical interactions
(Reipurth \& Clarke 2001), analogous to the formation of so-called run-away 
T~Tauri stars (Sterzik \& Durisen 1995, 1998; Durisen et al. 2001).
It is known that the dynamical evolution of gravitationally 
interacting systems of three or more bodies leads to frequent, close two-body
encounters and to the formation of close binary pairs 
out of the most massive objects in the system, 
as well as to the ejection of the lighter bodies into extended orbits or out of
the system with escape velocity (e.g. Valtonen \& Mikkola 1991).
The escape of the lightest body is an expected outcome, since the
escape probability scales approximately as the inverse third power of the mass.
The suggestion of this embryo-ejection model as the formation mechanism for brown dwarfs
has stimulated in past years
hydrodynamical collapse calculations (Bate et al. 2002, 2003; Bate \& Bonnell 2005)
and numerical N-body simulations 
(Sterzik \& Durisen 2003; Delgado-Donate et al. 2003, 2004; Umbreit et al. 2005)
which predict observable properties of brown dwarfs formed in this way.
There are significant differences in the theoretical approaches and predictions of these
models, which will be discussed later (see also Joergens 2005a for a recent review 
of brown dwarf formation).

An external process that prevents the stellar embryo from further accretion 
and growth in mass can also be photoevaporation by a strong UV wind from a nearby 
hot O or B star (Kroupa \& Bouvier 2003; Whitworth \& Zinnecker 2004).
Since we focus on brown dwarfs in Cha\,I, where there is no such hot star, 
we will not discuss this scenario further.

The ejection process might leave an observable imprint in the kinematics of 
members ejected from a cluster in comparison to that of non-ejected members.
In order to test this scenario, Joergens \& Guenther (2001) carried out a 
precise kinematic analysis of brown dwarfs in Cha\,I and 
compared it to that of T Tauri stars in the same field 
based on mean radial velocities (RVs) 
measured from high-resolution spectra taken with 
the UV-Visual Echelle Spectrograph (UVES) at the Very Large Telescope (VLT).
In this paper, we now present an improved analysis of this study based on 
additional RV measurements with UVES of several of the targets
in 2002 and 2004, based on a revised data analysis 
and on a cleaned and updated T~Tauri star sample as comparison.
Furthermore, the dispersion measured in terms of full width at half maximum (fwhm)
in Joergens \& Guenther (2001) was misinterpreted in the literature as standard deviation.
There is more than a factor of two difference between both quantities. In this paper,
we also give the dispersion in terms of the standard deviation and 
then discuss the implications.
We provide an empirical constraint for the kinematic properties of the studied group 
of very young
brown dwarfs in Cha\,I and discuss the results in the context of current ideas about the 
formation of brown dwarfs and, in particular, the theoretical predictions
of the embryo-ejection model.

It is noted that the UVES spectra were taken within the framework of an ongoing
RV survey for planetary and brown dwarf companions to young brown
dwarfs and very low-mass stars in Cha\,I, which was published elsewhere 
(Joergens 2003, 2005b).

The paper is organized as follows:
Sect.\,\ref{sect:spec} contains information about the UVES spectroscopy, 
data analysis, and the sample. 
In the next two sections, the kinematic study of brown dwarfs (Sect.\,\ref{sect:bds})
and of T~Tauri stars in Cha\,I (Sect.\,\ref{sect:tts}) is presented.
Section \ref{sect:discussion} contains a discussion of the results 
and a comparison with theoretical predictions,
and Sect.\,\ref{sect:concl} conclusions and a summary.

\section{UVES spectroscopy, RV determination, and sample} 
\label{sect:spec}

High-resolution spectra were taken for twelve brown dwarfs and
(very) low-mass stars in Cha\,I
between the years 2000 and 2004 with the echelle spectrograph UVES (Dekker\,et\,al.\,2000)
attached to the 8.2\,m Kueyen telescope of the VLT
operated by the European Southern Observatory at Paranal, Chile.
Details of the data acquisition, analysis, and RV determination are given in
Joergens (2005b).
However, we point out here that the errors given in Table 2 of Joergens \& Guenther (2001)
refer solely to \emph{relative} errors. Additionally,
an error of about 400\,m\,s$^{-1}$ due to the uncertainty in the zero offset of the template
has to be taken into account for the absolute RV.

The targets of these observations are brown dwarfs and (very) low-mass stars
with an age of a few million years
situated in the center of the nearby ($\sim$160\,pc) Cha\,I star-forming cloud
(Comer\'on et al. 1999, 2000; Neuh\"auser \& Comer\'on 1998, 1999).
Membership in the Cha\,I cluster, and therefore
the youth of the objects, is well established based on
H$\alpha$ emission, Lithium absorption, spectral types, and RVs
(references above; Joergens \& Guenther 2001; this work).
The measured mean RVs are listed in Table\,\ref{tab:bds} for the nine M6--M8 type
brown dwarfs and very low-mass stars Cha\,H$\alpha$\,1--8 and Cha\,H$\alpha$\,12,
while the mean RVs for B34 (M5), CHXR\,74 (M4.5) and Sz\,23 (M2.5)
are included in Table\,\ref{tab:tts} in the T~Tauri star list.

Deviations of RVs measured with UVES in this paper
compared to RVs presented in Joergens \& Guenther (2001) are
attributed to a spectroscopic companion for Cha\,H$\alpha$\,8 (Joergens 2005b)
and to improved data reduction for the other objects. 
For the latter, these deviations lie in the range of 3--80\,m\,s$^{-1}$ with the exception of
Cha\,H$\alpha$\,7 where the deviation was on the order of 3\,km\,s$^{-1}$, a situation that
can be attributed
to both the very low S/N of the spectra of this object and a
high sensitivity of the resulting RV on the extraction algorithm
for very low S/N spectra.

\section{Kinematics of brown dwarfs in Cha\,I}
\label{sect:bds}

The determined mean RVs for the nine M6 to M8-type
brown dwarfs and very low-mass stars
Cha\,H$\alpha$\,1--8 and Cha\,H$\alpha$\,12 are given in Table\,\ref{tab:bds}.
They range between 14.5 and 17.1\,km\,s$^{-1}$ with an arithmetic mean of 15.71$\pm$0.31\,km\,s$^{-1}$. 
The RV dispersion measured in terms of standard deviation of a population sample
is 0.92\,km\,s$^{-1}$ with an uncertainty of $\pm$0.32\,km\,s$^{-1}$
derived from error propagation.
The RV dispersion measured in terms of fwhm is 2.15\,km\,s$^{-1}$.
While the new mean RV is larger than the value of 14.9\,km\,s$^{-1}$
given by Joergens \& Guenther (2001),
their measurements of the total RV range (2.4\,km\,s$^{-1}$) and of the fwhm dispersion
(2.0\,km\,s$^{-1}$) differ only marginally from our results\footnote{The 
fwhm is related to the standard deviation
$\sigma$ of a Gaussian distribution by fwhm\,=\,$ \sigma \sqrt{8 \ln 2}$; however,
as pointed out by Sterzik \& Durisen (1998), the escape velocity distribution
might be significantly non-Gaussian.}.

The borderline between brown dwarfs and stars lies at about spectral type M7; i.e.
the sample in Table\,\ref{tab:bds} contains at least two objects,
Cha\,H$\alpha$\,4 and 5, which are most certainly of stellar nature.
The substellar border defined by the hydrogen-burning mass is a crucial dividing
line with respect to further evolution of an object, but there is no obvious reason
it should be of significance for the formation mechanism by which this object
was produced.
Thus, by whichever process brown dwarfs are formed, it is expected to work continuously
into the regime of very low-mass stars.
For an observational test to compare properties of brown dwarfs with
those of stars, it is hence not a priori clear where to set the
dividing line in mass of the two samples.

Therefore, we also consider the following samples:
a) a subsample of Table\,\ref{tab:bds}
containing only brown dwarfs and brown dwarf candidates, i.e. all
objects with spectral types M6.5 to M8 (7 objects); as well as
two larger samples that also 
include T~Tauri stars from Table\,\ref{tab:tts},
namely, b) a sample of brown dwarfs and (very) low-mass stars with spectral types M4.5 to M8
(11 objects); and c) a sample composed of all M-type (sub)stellar objects (M0--M8, 17 objects).
The mean RVs of these samples (15.87\,km\,s$^{-1}$, 15.77\,km\,s$^{-1}$, 15.36\,km\,s$^{-1}$) and
their RV dispersions (standard deviations: 0.97\,km\,s$^{-1}$, 0.83\,km\,s$^{-1}$, 1.20\,km\,s$^{-1}$)
do not differ significantly from
the values derived for the brown dwarf sample (M6--M8) that was initially chosen.
Also based on a Kolmogorov-Smirnov test, 
the RV distributions of the samples a) to c) are
consistent with the one of the M6-M8 sample
with a significance level of $\geq$ 99.98\% for a) and b)
and 98.50\% for c).

\begin{figure}[t]
\begin{center}
\includegraphics[height=.45\textwidth,angle=270,clip]{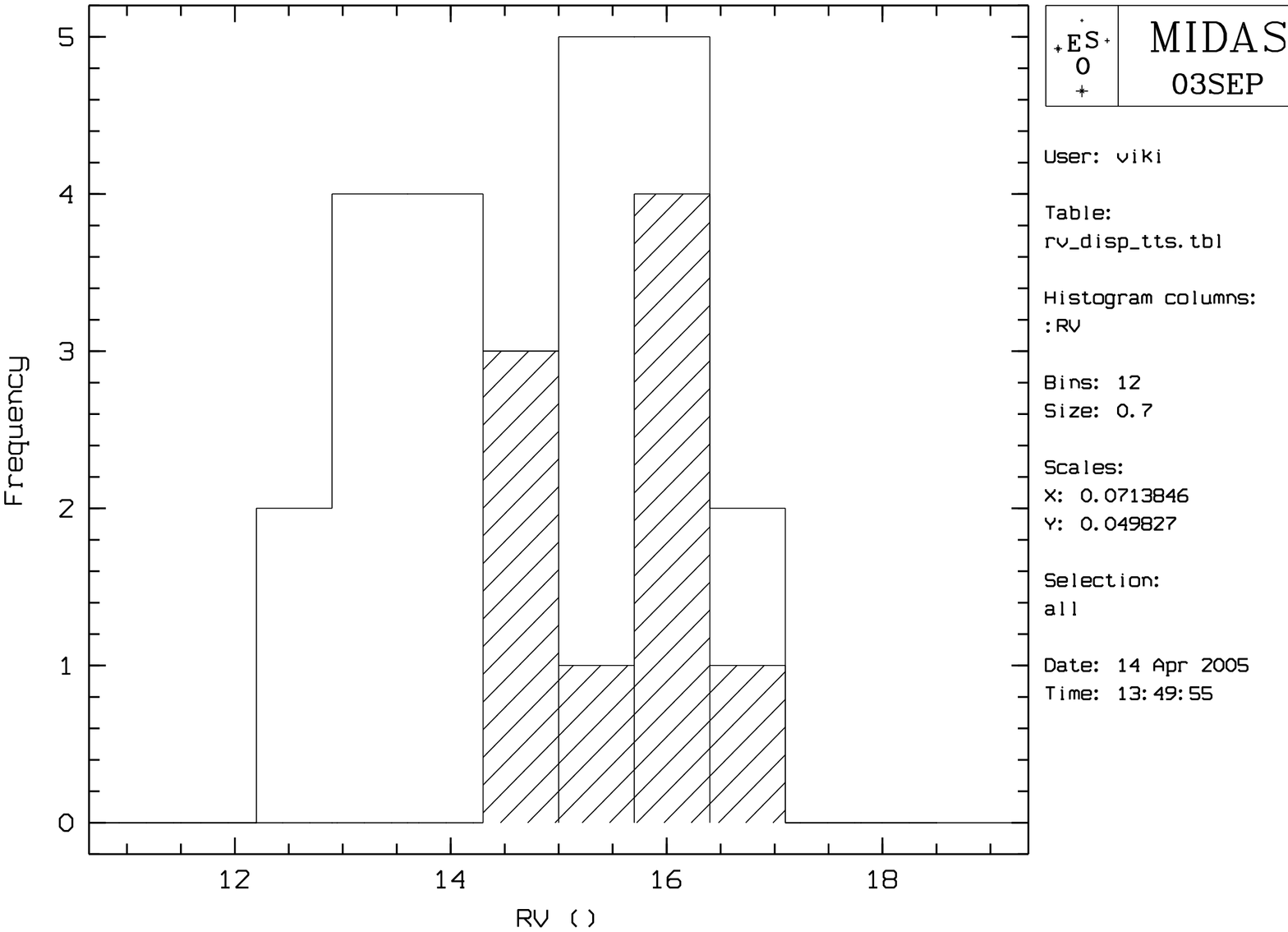}
\end{center}
\caption{
\label{hist}
\small{{\bf Histogram of mean RVs} in [km\,s$^{-1}$] of nine brown dwarfs
and very low-mass stars in Cha\,I ($\leq$ 0.1\,M$_{\odot}$, M6--M8, hashed)
and of 25 T~Tauri stars (M5--G2) in the same field.
}}
\end{figure}
%

\begin{figure}[t]
\begin{center}
\includegraphics[height=.45\textwidth,angle=90]{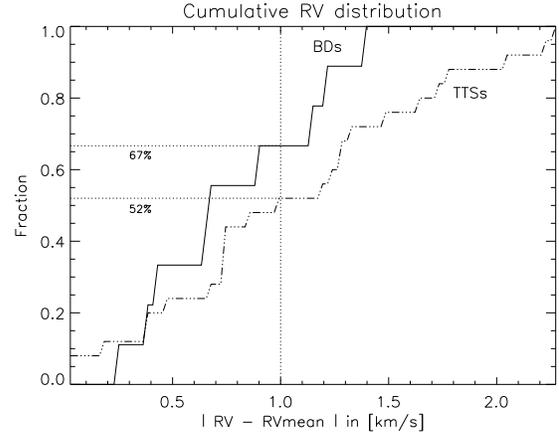}
\end{center}
\caption{
\label{cdf}
\small{{\bf Cumulative RV distribution} for
brown dwarfs (solid line) and T~Tauri stars (dash-dotted line).
The samples are the same as for Fig.\,\ref{hist}.
Plotted is the fraction of objects with a relative RV smaller or equal to a given relative RV.
It can be seen that the RV of none of the brown dwarfs 
deviates from the mean RV of the brown dwarfs by more than 1.4\,km\,s$^{-1}$ and that
67\% of them have RV deviations smaller or equal to 0.9\,km\,s$^{-1}$
compared to 52\% for the T~Tauri stars with RV deviations smaller or equal to 1.0\,km\,s$^{-1}$.
}}
\end{figure}
%

\begin{figure}[t]
\begin{center}
\includegraphics[height=.45\textwidth,angle=90]{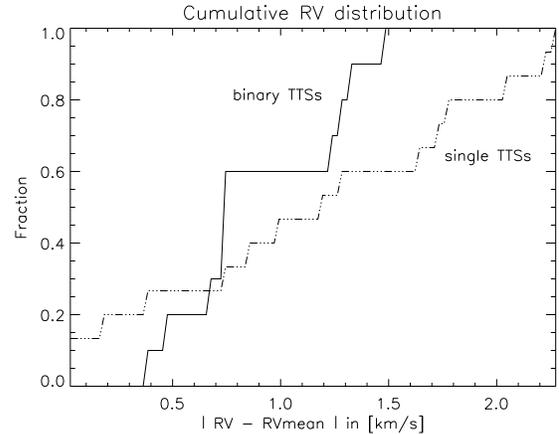}
\end{center}
\caption{
\label{fig:cdf_tts}
\small{{\bf Cumulative RV distributions} for T~Tauri `singles' (dash-dotted line) and 
binaries\,/\,multiples (solid line). Note the steeper rise of the binary subsample 
indicating an absence of a high velocity tail compared to those T~Tauri stars regarded as single. 
}}
\end{figure}
%

\begin{figure}[t]
\begin{center}
\includegraphics[height=.45\textwidth,angle=90]{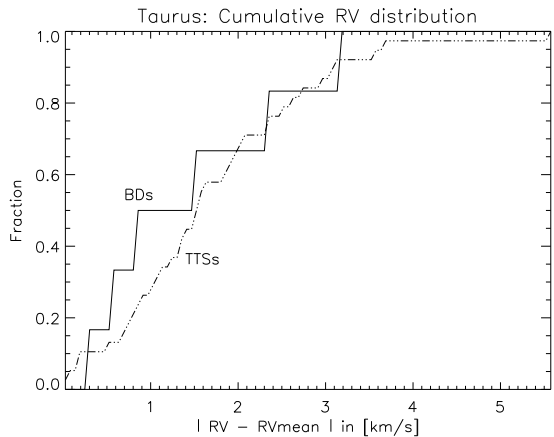}
\end{center}
\caption{
\label{cdf_taurus}
\small{{\bf For comparison, cumulative RV distributions for members of Taurus:}
brown dwarfs, very low-mass stars 
(M6 and later, six objects, White \& Basri 2003, solid line), and higher mass
T~Tauri stars (38 objects, Hartmann et al. 1986, dash-dotted line).
}}
\end{figure}
%

\section{Kinematics of T~Tauri stars in Cha\,I}
\label{sect:tts}

In order to compare the kinematics of the brown dwarfs in Cha\,I with
higher-mass stellar objects in this cluster,
we compiled all T~Tauri stars that are confined to the same region 
for which RVs have been measured with a precision of 2\,km\,s$^{-1}$ or better
from the literature 
(Walter 1992; Dubath et al. 1996; Covino et al. 1997; Neuh\"auser \& Comer\'on 1999), 
from Guenther et al. (in prep., see Joergens \& Guenther 2001),
and from our own measurements based on UVES spectra for mid- to late M-type Cha\,I members.
The result is a sample of 25 T~Tauri stars (spectral types M5--G2)
as listed in Table\,\ref{tab:tts}.
For nine of them, RV measurements are available from more than one author
and for two objects, CHX18N (Walter et al. 1992; Covino et al. 1997) and CHXR74
(Joergens 2005b), RV measurements at different epochs are significantly discrepant, which
hints at long-period spectroscopic companions.
Table\,\ref{tab:tts} gives the
RVs measured by the different authors as well as the derived mean RV that
was adopted for this study in the last column.

Compared to Joergens \& Guenther (2001),
the T~Tauri sample was revised by identifying
double entries in their Table\,2, and by taking 
additional RVs from the literature into account, by rejection of two
foreground objects in the previous sample, by an improved
data reduction for UVES-based RVs, as well as by additional UVES
measurements. Details are given in the appendix Sect.\,\ref{sect:app}.

We found that these 25 T~Tauri stars
have an arithmetic mean RV of 14.73$\pm$0.25\,km\,s$^{-1}$, and
an RV dispersion in terms of a standard deviation of
1.26$\pm$0.31\,km\,s$^{-1}$ and in terms of an fwhm of 2.96\,km\,s$^{-1}$.
Compared to the values given by Joergens \& Guenther (2001),
namely a mean RV of 14.9\,km\,s$^{-1}$, a standard deviation of 1.5\,km\,s$^{-1}$, and an
fwhm of 3.6\,km\,s$^{-1}$, the new values are slightly lower.
Interestingly, the difference in the dispersion can be solely attributed to the previously
unresolved binarity of CHX18N, and the difference in the mean RV can be partly attributed to
this fact (cf. Sect.\,\ref{sect:app}).

As for the brown dwarf case, we also calculated the kinematics for subsamples 
in order to account for the possibility that, if brown dwarfs are formed by a 
different mechanism than stars, (very) low-mass stars might form in a brown dwarf like 
manner rather than a star like.
No significant differences were found 
between the original T~Tauri sample and a sample of 
only those stars with a larger mass than about 0.2\,M$_{\odot}$
(23 stars, M3.25--G2, mean RV: 14.61\,km\,s$^{-1}$, RV standard deviation: 1.24\,km\,s$^{-1}$) 
and a sample of only K and G type T~Tauri stars 
(17 stars, K8--G2, mean RV: 14.62\,km\,s$^{-1}$, RV standard deviation: 1.21\,km\,s$^{-1}$ ).
A Kolmogorov-Smirnov test also showed that the RV distributions of the two subsamples
are consistent with the one of the original T~Tauri sample
with significance levels above 99.52\%.

A kinematic difference between single and binary\,/\,multiple stars was suggested
by Sterzik \& Durisen (2003) and Delgado-Donate et al. (2003) in the sense that
the velocities of multiples are less dispersed.
Therefore, we investigate the kinematics of the T~Tauri stars in respect
of their multiplicity status.
A significant fraction of the T~Tauri stars in our sample has been resolved into visual
binaries (Sz\,19, Sz\,20, Sz\,41, F\,34, CV\,Cha, SX\,Cha, VW\,Cha, CHX\,22)
by Reipurth \& Zinnecker (1993), Brandner et al. (1996),
Brandner \& Zinnecker (1997), and Ghez et al. (1997).
Furthermore, as mentioned above, there are indications of spectroscopic companions around
CHX18N (Walter 1992; Covino et al. 1997) and CHXR\,74 (Joergens 2005b).
Thus, among the sample of 25 T~Tauri stars, at least about 10 are in binary or multiple systems.
We calculated a mean RV of 14.68\,km\,s$^{-1}$ and an RV dispersion (standard deviation) of
1.02\,km\,s$^{-1}$
for the sample of the 10 T~Tauri stars in binary or multiple systems and
a mean RV of 14.76\,km\,s$^{-1}$ and a RV dispersion (standard deviation) of 1.42\,km\,s$^{-1}$ for
the remaining stars of the sample, which might, for the most part, be single.
Thus, the RVs of the sample of `single' T~Tauri stars in Cha\,I are slightly more dispersed than
the T~Tauri binary stars, which are mainly visual and, hence, wide systems,
as further discussed in the next section.

\begin{table}[t]
\caption{
\label{tab:bds}
Mean RVs for brown dwarfs and very low-mass stars in Cha\,I measured from UVES spectra. 
The values are the weighted mean of the individual time--resolved RVs
from Joergens (2005b) and the error of the weighted mean plus 400\,m\,s$^{-1}$
for the error in the zero point of the velocity.
For Cha\,H$\alpha$\,8, the RV obtained in 2002 deviates significantly from the one 
obtained in 2000, hinting at a spectroscopic binary; therefore, the listed RV of Cha\,H$\alpha$\,8 is
the arithmetic mean of the separately calculated 2000 and 2002 mean values.
}
\vspace{0.1cm}
\begin{tabular}{lcc}
\hline
\hline
\myrule
object              & SpT   & RV$_\mathrm{UVES}$ \\
                    &       & [km\,s$^{-1}$]     \\
\hline
\hline
\myrule
Cha\,H$\alpha$\,1   & M7.5  & 16.35 $\pm$ 0.63 \\
Cha\,H$\alpha$\,2   & M6.5  & 16.13 $\pm$ 0.53 \\
Cha\,H$\alpha$\,3   & M7    & 14.56 $\pm$ 0.60 \\
Cha\,H$\alpha$\,4   & M6    & 14.82 $\pm$ 0.40 \\
Cha\,H$\alpha$\,5   & M6    & 15.47 $\pm$ 0.43 \\
Cha\,H$\alpha$\,6   & M7    & 16.37 $\pm$ 0.68 \\
Cha\,H$\alpha$\,7   & M8    & 17.09 $\pm$ 0.98 \\
Cha\,H$\alpha$\,8   & M6.5  & 16.08 $\pm$ 1.62 \\
Cha\,H$\alpha$\,12  & M7    & 14.50 $\pm$ 0.96 \\
\hline
\hline
\vspace{0.5cm}
\end{tabular}
\end{table}

\begin{table*}[t]
\caption{
\label{tab:tts}
List of T~Tauri stars in Cha\,I with known RVs ordered by spectral type.
All T~Tauri stars in Cha\,I were
compiled with RVs known to a precision of 
2\,km\,s$^{-1}$ or better from our UVES observations and from the literature.
RV$_\mathrm{UVES}$ values are measured from our UVES spectra and are weighted means based on 
several measurements, and the error takes into account 400\,m\,s$^{-1}$
for the error in the zero point of the velocity.
The last column gives the RV adopted for the kinematic study.
For objects with more than one available RV measurement,
usually the weighted average and the error of the weighted mean are given.
An exception of this was made for CHX18N and CHXR\,74,
for which RVs at different epochs are significantly deviant; and, therefore,
the arithmetic mean of the different epoch RVs is given. 
The listed names are those used by the authors of the RV data.
In addition, for each object we give the name used by the most recent
compilation of Cha\,I members (Luhman 2004), e.g. `T5' stands for the Simbad name `Ass Cha T 2-5'.
Errors for CS\,Cha, VZ\,Cha, WY\,Cha are not given by the author and are rough estimates
based on an average error for the remaining RVs determined by the same author.
References are W92: Walter 1992;
D96: Dubath et al. 1996;
C97: Covino et al. 1997;
N99: Neuh\"auser \& Comer\'on 1999;
G: Guenther et al.,in prep.;
UVES: Joergens \& Guenther 2001, Joergens 2003, Joergens 2005b, this work.
}
\vspace{0.1cm}
\begin{footnotesize}
\begin{tabular}{llcccccccc}
\hline
\hline
\myrule
object   & other names               & SpT   &RV$_{\rm W92}$&RV$_{\rm D96}$&RV$_{\rm C97}$&RV$_{\rm N99}$&RV$_{\rm G}$  & RV$_{\rm UVES}$ &RV$_{\rm mean}$ \\
         &                           &       &[km\,s$^{-1}$]&[km\,s$^{-1}$]&[km\,s$^{-1}$]&[km\,s$^{-1}$]&[km\,s$^{-1}$]& [km\,s$^{-1}$]  & [km\,s$^{-1}$] \\
\hline
\hline
\myrule
Sz\,19   & \tiny{DI\,Cha, CHX9, T26} & G2    & 14$\pm$2 & 13.5$\pm$0.6 &              &              &              &                 & 13.5$\pm$0.1 \\
CHX\,7   & \tiny{T21}                & G5    & 17$\pm$2 &              &              &              &              &                 & 17$\pm$2     \\
CHX\,22  & \tiny{T54}                & G8    & 14$\pm$2 &              &              &              &              &                 & 14$\pm$2     \\
CV Cha   & \tiny{Sz\,42, T52}        & G9    &          & 15.1$\pm$0.3 &              &              & 15.6$\pm$0.9 &                 & 15.2$\pm$0.2 \\
Sz\,6    & \tiny{CR\,Cha, T8}        & K2    &          & 14.9$\pm$0.8 &              &              &              &                 & 14.9$\pm$0.8 \\
F\,34    & \tiny{WKK F34, CHXR47}    & K3    &          &              & 14.0$\pm$2.0 &              &              &                 & 14.0$\pm$2.0 \\
Sz\,41   & \tiny{T51,RXJ1112.7-7637} & K3.5  &          & 13.9$\pm$0.4 & 16$\pm$2     &              &              &                 & 14.0$\pm$0.4 \\
CHX20E   & \tiny{CHXR55}             & K4.5  & 13$\pm$2 &              &              &              &              &                 &  13$\pm$2    \\
CT Cha   & \tiny{Sz\,11, T14}        & K5    &          & 15.1$\pm$0.5 &              &              & 15.5$\pm$1.4 &                 & 15.1$\pm$0.1 \\
CHX18N   & \tiny{RXJ1111.7-7620}     & K6    & 13$\pm$2 &              & 19.0$\pm$2.0 &              &              &                 & 16.0$\pm$3.0 \\
CHX\,10a & \tiny{CHX\,10, CHXR\,28}  & K6    & 16$\pm$2 &              &              &              &              &                 & 16$\pm$2     \\
VZ Cha   & \tiny{T40}                & K6    &          &              &              &              & 14.7$\pm$0.8 &                 & 14.7$\pm$0.8 \\
CS Cha   & \tiny{Sz\,9, T11}         & K6    &          & 14.7$\pm$0.3 &              &              & 14.9$\pm$0.8 &                 & 14.7$\pm$0.1 \\
CHXR\,37 & \tiny{RXJ1109.4-7627}     & K7    &          &              & 13.1$\pm$2.0 &              &              &                 & 13.1$\pm$2.0 \\
TW Cha   & \tiny{T7}                 & K8    &          &              &              &              & 15.7$\pm$1.2 &                 & 15.7$\pm$1.2 \\
VW Cha   & \tiny{T31}                & K8    &          &              &              &              & 15.1$\pm$0.1 &                 & 15.1$\pm$0.1 \\
WY Cha   & \tiny{Sz\,36, T46}        & K8    &          & 12.9$\pm$0.9 &              &              & 12.1$\pm$0.8 &                 & 12.5$\pm$0.4 \\
SX Cha   & \tiny{T3}                 & M0.5  &          &              &              &              & 13.4$\pm$0.9 &                 & 13.4$\pm$0.9 \\
SY Cha   & \tiny{T4}                 & M0.5  &          &              &              &              & 12.7$\pm$0.1 &                 & 12.7$\pm$0.1 \\
CHX\,21a & \tiny{CHX\,21, CHXR\,54}  & M1    & 14$\pm$2 &              &              &              &              &                 & 14$\pm$2     \\
Sz\,20   & \tiny{VV\,Cha, T27}       & M1    &          & 15.4$\pm$1.3 &              &              &              &                 & 15.4$\pm$1.3 \\
Sz\,23   & \tiny{T30}                & M2.5  &          &              &              &              &              &15.57$\pm$0.55   & 15.57$\pm$0.55\\
Sz\,4    & \tiny{T5}                 & M3.25 &          & 16.5$\pm$1.3 &              &              &              &                 & 16.5$\pm$1.3 \\
CHXR\,74 &                           & M4.5  &          &              &              & 16.5$\pm$1.0 &              &14.58$\pm$0.62   &               \\
         &                           &       &          &              &              &              &              &17.42$\pm$0.44   & 16.2$\pm$0.8  \\
B\,34    & \tiny{CHXR\,76}           & M5    &          &              &              & 17.6$\pm$1.7 &              &15.75$\pm$0.42   & 15.9$\pm$0.4\\
%
\hline
\hline
\vspace{1cm}
\end{tabular}
\end{footnotesize}
\end{table*}

\section{Discussion}    
\label{sect:discussion}

\subsection{Observational results for Cha\,I}

The mean RV of the M6--M8 type brown dwarfs and very low-mass stars 
(15.7$\pm$0.3\,km\,s$^{-1}$) is consistent with that of the surrounding molecular gas  
(15.4\,km\,s$^{-1}$, Mizuno et al. 1999) in agreement with membership of this substellar population 
in the Cha\,I star-forming cloud.
In Fig.\,\ref{hist} the RVs determined for the brown dwarfs in Cha\,I are compared 
to those of T~Tauri stars in the same region in the form of a histogram.
It can be seen that the RVs of the brown dwarfs are on average 
larger, by 1.8 times the errors,
than those for the T~Tauri stars (14.7$\pm$0.3\,km\,s$^{-1}$).
The dispersion of the RVs of the brown dwarfs measured in terms of standard deviation  
(0.9$\pm$0.3\,km\,s$^{-1}$) is slightly lower than that of the
T~Tauri stars (1.3$\pm$0.3\,km\,s$^{-1}$), 
and both are (slightly) larger than that of the surrounding molecular gas 
(the fwhm of 0.9\,km\,s$^{-1}$ given by Mizuno et al. (1999) for the
average cloud core translates to a standard deviation of 0.4\,km\,s$^{-1}$).
While the differences in the dispersion values lie well within the error range,
it is unlikely from the data that the RVs of the brown dwarfs are
significantly more dispersed than that of the T~Tauri stars.

Since the velocity distribution arising from the dynamical decay of N-body clusters
might be significantly non-Gaussian (Sterzik \& Durisen 1998),
the determined RVs of the brown dwarfs and T~Tauri stars
in Cha\,I are also analyzed in the form of cumulative distributions.
In Fig.\,\ref{cdf}, the fraction of objects with a relative RV smaller or equal to a given
relative RV is plotted. The relative RV is given by the absolute value of the
deviation from the mean of the whole group.
Since the measured mean RVs for the brown dwarfs and T~Tauri stars deviate slightly but significantly,
these two different mean RVs are used.
It can be seen that the cumulative RV distributions for brown dwarfs and T~Tauri stars
in Cha\,I diverge for higher velocities.
While the cumulative RV distribution of the brown dwarfs displays a linear and steeper
increase,
with none of the them having an RV deviating from the mean by more than 1.4\,km\,s$^{-1}$,
the T~Tauri stars show a tail of higher velocities: 28\% have an RV deviating from the
mean by more than 1.4\,km\,s$^{-1}$.
An examination of the RV distributions of the studied brown dwarf and T~Tauri star sample in Cha\,I
based on Kolmogorov-Smirnov statistics also indicates a significant
difference. The significance level for the consistency of the distributions of the absolute RVs
is only 11.70\%, while it is 74.30\% for the relative RVs considered in the same manner as for Fig.\,\ref{cdf}.

That the improved value for the RV dispersion in terms of standard deviation
of the T~Tauri stars is slightly smaller than
the value found by Joergens \& Guenther (2001)
can be attributed solely to the unresolved binarity of one of the T~Tauri stars in the sample.
This demonstrates that the orbital motion of unresolved spectroscopic binaries are 
a source of error.
An additional possible error source is RV variability induced by surface spots.
Given the recent indications that RV noise caused by surface activity is very small for brown dwarfs
in Cha\,I (Joergens 2005b),
the slightly higher value for the RV dispersion of T~Tauri stars compared to brown dwarfs
might be attributed to such systematic errors being more pronounced in the stellar than in the substellar
regime.
While the RVs of the T~Tauri stars have to a large degree been
collected from the literature and obtained with a variety of instruments and wavelength
resolutions, there is no hint that this introduces a systematic error. 
First, for seven out of the nine T~Tauri stars for which RVs have been measured by more than one author,
consistent RVs were found despite different instrumentation. Secondly, RV dispersions and mean values
of the T~Tauri subsamples in Table\,\ref{tab:tts} 
sorted by author, i.e. obtained in a uniform way, agree very well with the kinematic properties 
derived here for the total sample of compiled T~Tauri stars. For example,
the T~Tauri stars observed by Guenther et al. have an RV dispersion of 1.3\,km\,s$^{-1}$
and a mean value of 14.4\,km\,s$^{-1}$; the ones observed by Dubath et al. (1996) 
have 1.1\,km\,s$^{-1}$ and 14.7\,km\,s$^{-1}$; and the ones observed by Walter (1992) have
1.5\,km\,s$^{-1}$ and 14.4\,km\,s$^{-1}$.

An investigation of the RVs of the T~Tauri stars in Cha\,I taking their multiplicity status
into account showed that the sample of 10 stars for which indications of visual or spectroscopic binaries
were detected has less dispersed RVs (standard deviation 1.02$\pm$0.57\,km\,s$^{-1}$)
compared to the remaining 15 `single' stars of the sample (1.42$\pm$0.41\,km\,s$^{-1}$),
while these values are still consistent within the errors.
In Fig.\,\ref{fig:cdf_tts}, cumulative RV distributions
for T~Tauri binaries and `singles' are plotted. A trend of less dispersed
velocities for the binaries can be seen, with 60\% of the T~Tauri `singles' 
having v$\leq$1.5\,km\,s$^{-1}$ but
100\% of T~Tauri binaries having v$<$1.5\,km\,s$^{-1}$.
A Kolmogorov-Smirnov test also indicates that the RV distributions of these samples might
be different (71.59\%).
A similar study for the brown dwarfs in Cha\,I 
is not possible yet since among
the brown dwarf sample, so far only one 
has shown any indication of binarity (Cha\,H$\alpha$\,8, Joergens 2005b). 
The fact that this object has one of the smallest deviations
from the mean of the whole group, namely 0.4\,km\,s$^{-1}$, is in line with  
observations of the T~Tauri binaries.   
However, several of the (sub)stellar objects regarded as `single' 
may still be resolved into multiple systems in the future.
Since nine out of these ten T~Tauri binaries are \emph{visual}, hence, wide binaries,
we can assume that predominantly spectroscopic or, at least, close systems have still not been resolved.
Thus, this kinematic difference might translate into a kinematic difference between
the group of wide binaries and the group of single stars and close binaries.

\subsection{Comparing Cha\,I and Taurus observations}
\label{sect:chaTau}

Comparing our kinematic study in Cha\,I with RV measurements for the Taurus star-forming region shows 
that, 
in agreement with our finding for Cha\,I, the RV dispersion for brown dwarfs in Taurus does not
seem to deviate 
significantly from that of T~Tauri stars in this cloud.
The RV dispersion measured in terms of the standard deviation of six
brown dwarfs and very low-mass stars in Taurus (M6--M7.5) 
is 1.9\,km\,s$^{-1}$ (White \& Basri 2003) and that for 38 T~Tauri stars 
in the same cloud is 2.1\,km\,s$^{-1}$ (Hartmann et al. 1986).

There are still two differences compared to the situation in Cha\,I.
In Fig.\,\ref{cdf_taurus}, the cumulative RV distributions are plotted 
for both Taurus samples based on the RVs published
by the authors. Both follow the same distribution in this diagram,
in contrast to the case for Cha\,I (Fig.\,\ref{cdf}) where
the cumulative RV distributions for brown dwarfs and stars diverge for higher velocities.

Secondly, the RV dispersions for Taurus brown dwarfs and stars are significantly higher than
the ones for Cha\,I brown dwarfs (0.9\,km\,s$^{-1}$) and stars (1.3\,km\,s$^{-1}$).
We measured a global RV dispersion of 1.24$\pm$0.24\,km\,s$^{-1}$
for all Cha\,I brown dwarfs and stars in 
Tables\,\ref{tab:bds} and \ref{tab:tts}.
For comparison, the velocity dispersion of the molecular gas in the Cha\,I cloud cores
is on average 0.4\,km\,s$^{-1}$ in terms of the standard deviation
(Mizuno et al. 1999).
For Taurus, we found\footnote{For the calculation of the error of the dispersion, the individual
errors of the RV values are necessary, which are not always given by Hartmann et al. (1986).
In the cases of absent individual errors, we used an average error derived from the given individual errors.} a global RV dispersion of 2.04$\pm$0.30\,km\,s$^{-1}$
for the sample of brown dwarfs and (very) low-mass stars  
in White \& Basri (2003) combined with the sample of T~Tauri stars in Hartmann et al. (1986).
Again for comparison, the velocity dispersion of the molecular gas in the Taurus
cloud cores is on average 0.3\,km\,s$^{-1}$ in terms of standard deviation
(Onishi et al. 1996).
We conclude that the RV dispersion for Cha\,I and Taurus (sub)stellar members deviate
significantly. The RV dispersion for Taurus is about a factor of two higher than for Cha\,I,
while Taurus has a much lower stellar density and star-formation 
efficiency (Oasa et al. 1999; Tachihara et al. 2002).
Thus, a fundamental increase in velocity dispersion with stellar density of the star-forming region,
as suggested by Bate \& Bonnell (2005), is not established observationally
(see also Sect.\,\ref{sect:sph}).

\subsection{Comparison with theoretical models}

\subsubsection{Overview of models}

Enormous theoretical efforts have been undertaken in recent years to model the formation
of brown dwarfs by the embryo-ejection mechanism and to simulate the dynamical evolution in
star-forming regions. Hydrodynamical calculations of the collapse of a 50\,M$_{\odot}$ cloud
(Bate et al. 2002, 2003; Bate \& Bonnell 2005) have demonstrated that brown dwarfs are formed 
in these models as stellar seeds that are ejected early.
While predictions of (sub)stellar parameters by these models are based on small numbers,
this can be overcome by the combination of collapse calculations with
N-body simulations of further dynamical evolution 
(Delgado-Donate et al. 2003, 2004). 
However, the predicted properties of current hydrodynamical models have been questioned 
because of the lack of feedback processes (Kroupa \& Bouvier 2003a).

On the other hand, simulations of the dynamical evolution of small N-body clusters (Sterzik \& Durisen 2003)
made statistically robust predictions of the properties of very young brown dwarfs and stars possible. 
There have been recent efforts to incorporate further details of the star-formation process, 
in particular ongoing accretion during the dynamical
interactions (Umbreit et al. 2005).
However, the current N-body simulations do not consider the gravitational potential of the cluster, 
which might significantly influence, e.g., the predicted velocities.

An approach to estimating the gravitational potential of existing star-forming regions has been
suggested by Kroupa \& Bouvier (2003b) 
for Taurus-Auriga and the Orion Nebula Cluster (ONC). 
However, the assumed cloud core properties are not straightforward to understand. While 
the cluster mass (stars and gas) of 9000\,M$_{\odot}$ adopted by the authors for the ONC is plausible
for the embedded gas of the whole ONC at birth time (Wilson et al. 2005),
the value of 50\,M$_{\odot}$ for Taurus-Auriga corresponds to about twice the mass of \emph{one}
C18O cloud core in Taurus (Onishi et al. 1996). Furthermore, 
since Cha\,I cloud cores have a similar average mass and radius (Mizuno et al. 1999) to Taurus,
we would find a similar estimate for the 
gravitational potential when following along these lines,
while there is a large difference in stellar density and star-formation efficiency
between these regions (Oasa et al. 1999; Tachihara et al. 2002).

\subsubsection{RVs and 3D velocities in Cha\,I}

When considering dynamical evolutions,
the transformation from RVs to 3D velocities is not
straightforward, but instead depends on details of the gravitational potential of the
cluster and on the number of objects with very small velocities, i.e. the details of the
simulations, as explained in the following. The brown dwarfs studied in Cha\,I 
are situated in a relatively densely populated region of Cha\,I
at the periphery of one of its six cloud cores, the so-called `YSO condensation B'
(Mizuno et al. 1999). They occupy a field of less than 0.2$\times 0.2 \deg$
at a distance of 160\,pc. Having an age of about 2\,Myr (Comer\'on et al. 2000),
the brown dwarfs born within this field and ejected during their formation in
directions with a significant fraction perpendicular to the line-of-sight, would
have already vanished for velocities of 0.4\,km\,s$^{-1}$ or larger
(Joergens et al. 2003a).
With a velocity of $\sim$0.8\,km\,s$^{-1}$, an object can even cross the whole extent of the
YSO condensation B ($\sim$0.2$\times$0.5$\deg$).
In the case of a significant fraction of brown dwarfs leaving the survey area,
the remaining observable objects will be of two sorts: those with 
very small velocities (too small to travel the 
extent of the region ($<$0.4\,km\,s$^{-1}$) and/or 
too small to overcome the binding energy of the cluster)  
and those with larger velocities but moving predominantly in a radial 
direction.
Due to this selection of fast-moving objects predominantly in a radial direction, 
the observed RV dispersion
of such a group in such a limited survey area
would be larger than the calculated 1D velocities that consider all objects regardless of the distance 
to the birth place.
On the other hand, the observed RV dispersion would still be smaller than the calculated 3D velocities
given the majority of bound objects with a small velocity component.

\subsubsection{Comparison with hydrodynamical calculations}
\label{sect:sph}

The brown dwarfs and stars formed in the hydrodynamical model by Bate et al. (2003) 
share the same kinematic properties.
For a stellar density of 1.8~10$^3$\,stars/pc$^3$, they
find an RMS velocity dispersion of 2.1\,km\,s$^{-1}$ in 3D.
Identical calculations for a denser star-forming region (2.6~10$^4$\,stars/pc$^3$, 
Bate \& Bonnell 2005) yield an RMS velocity dispersion of 4.3\,km\,s$^{-1}$.
The combined hydrodynamic / N-body model by 
Delgado-Donate et al. (2004) 
also predict no, or at most slight kinematic
differences between ejected and non-ejected members of a cluster.
The 3D velocity dispersion of the produced objects is
2--3\,km\,s$^{-1}$ (Delgado-Donate, pers.comm.) for modeled
stellar densities of 2.5--3.6~10$^4$ stars/pc$^{3}$ (Delgado-Donate et al. 2004).

The theoretical finding that there is no kinematic difference between stars and brown dwarfs is consistent with
our measurement of no significant difference between the RV dispersion of brown dwarfs 
(0.9$\pm$0.3\,km\,s$^{-1}$) and stars (1.3$\pm$0.3\,km\,s$^{-1}$) in Cha\,I.
However,
the RV dispersions observed in Cha\,I for both brown dwarfs and stars 
are much smaller than predicted by the models.
Since the stellar densities in these calculations (on the order of 10$^{3}$ to
10$^{4}$\,stars/pc$^3$)
are much higher than in the Cha\,I cloud (on the order of 10$^{2}$\,stars/pc$^3$,
Oasa et al. 1999),
one could argue that extrapolation towards smaller densities
might explain the discrepancy.
An extrapolation of the trend seen in the models by Bate et al. (2003) and Bate \& Bonnell (2005)
towards the stellar densities in Cha\,I, for instance, would yield a velocity dispersion of
about 1\,km\,s$^{-1}$ that is consistent with the RV dispersion measured for Cha\,I. 
However, the same extrapolation yields a value that is highly inconsistent with the 
observations for Taurus that has a much smaller observed stellar density than Cha\,I
(see Sect.\,\ref{sect:chaTau}).
Furthermore, there are apparently inconsistencies between different theoretical models:
the model from Delgado-Donate et al. (2004) 
produces similar stellar densities as calculation~2 of Bate \& Bonnell (2005), but
at the same time both predict very different velocity dispersions
of 2--3\,km\,s$^{-1}$ (Delgado-Donate, pers.comm.) and  4.3\,km\,s$^{-1}$ (Bate \& Bonnell 2005).
We therefore conclude that the dependence of the velocity dispersion on the stellar
densities 
has not yet been established and an extrapolation is not advisable.

\subsubsection{Comparison with N-body simulations}

The decay models of Sterzik \& Durisen (2003) predict that 25\% of the
brown dwarf singles have a velocity that is smaller than 1\,km\,s$^{-1}$. This is a smaller
percentage than found by our observations that 
67\% of the brown dwarfs in Cha\,I have RVs smaller than 1\,km\,s$^{-1}$.
Furthermore, Sterzik \& Durisen (2003) find a high-velocity tail with
40\% single brown dwarfs having v$>$1.4\,km\,s$^{-1}$ and 10\% having v$>$5\,km\,s$^{-1}$.
This is also not seen in our data, where none has an RV deviating by the mean RV of the group by more than
1.4\,km\,s$^{-1}$. Admittedly, the relatively small size of our brown dwarf sample does not exclude 
the possibility that
we have missed the 40\% objects moving faster than 1.4\,km\,s$^{-1}$.

In the formation phase where dynamical interactions become important,
gas accretion might still be a significant factor.
N-body simulations by Umbreit et al. (2005) find
higher ejection velocities for models  
taking ongoing accretion during the dynamical encounters into account.
They predict that between 60\% and almost 80\% of single brown dwarfs have
velocities larger than 1\,km\,s$^{-1}$ depending on the accretion rates.
That is much larger than found by our observations, where only about 30\% of the brown dwarfs have
velocities $>$ 1\,km\,s$^{-1}$.

Sterzik \& Durisen (2003) furthermore
find different kinematics for singles and binaries:
90\% of their stellar binaries have velocities smaller than 1\,km\,s$^{-1}$, while
only 50\% of stellar singles can be found in that velocity range.
This agrees with our tentative finding that the subgroup of binaries among the studied 
sample of T~Tauri stars
in Cha\,I has a lower RV dispersion (1.0\,km\,s$^{-1}$) and no high-velocity tail
(Fig.\,\ref{fig:cdf_tts}) compared to the remaining `single' stars (1.42\,km\,s$^{-1}$).
However, the T~Tauri `single' star sample might be contaminated by unresolved binaries.

\section{Conclusions and summary}
\label{sect:concl}

In order to pave the way to an understanding of the still unknown origins of brown dwarfs,
we explored the kinematic properties of extremely young brown dwarfs in the star-forming
cloud Cha\,I based on precise RVs measured from high-resolution UVES\,/\,VLT spectra.
This kinematic study is an improved version of the one by Joergens \& Guenther (2001),
which provided the first
observational constraints for the velocity distribution of a group of very young 
brown dwarfs.

We found that nine brown dwarfs and very low-mass stars (M6--M8, M\,$\la$0.1\,M$_{\odot}$)
in Cha\,I kinematically form a very homogeneous group.
They have very similar absolute RVs with a mean value of 15.7\,km\,s$^{-1}$, an 
RV dispersion in terms of standard deviation of 0.9\,km\,s$^{-1}$, and a total covered RV range of 2.6\,km\,s$^{-1}$.

We conducted a comparison of the kinematic properties of these brown dwarfs with those
of 25 T~Tauri stars confined to the same field 
based on our UVES measurements, as well as on RVs from the literature.
For the T~Tauri stars, we
determined a mean RV of 14.7\,km\,s$^{-1}$, a RV dispersion of 1.3\,km\,s$^{-1}$, and a total
range of 4.5\,km\,s$^{-1}$.

The mean RVs of the brown dwarfs are larger than that of the T~Tauri stars
by less than two times the errors; however, both values are consistent
with the velocity of the molecular gas of the surroundings.
The RV dispersion measured for the brown dwarfs is slightly smaller than the one for the T~Tauri stars,
but this difference lies within the errors.
We found that the cumulative RV distributions for the brown dwarfs and for the T~Tauri stars diverge for
RVs higher than about 1\,km\,s$^{-1}$, with the brown dwarfs displaying no tail with high velocities in 
contrast to the 
T~Tauri stars. This could be an intrinsic feature or might be attributed to more pronounced systematic 
RV errors in the stellar
than in the substellar mass domain.

The finding of consistent RV dispersions for brown dwarfs and stars in Cha\,I
(Joergens \& Guenther 2001; this paper) is also seen in RV data for brown dwarfs and stars in the Taurus
star-forming region (Hartmann et al. 1986; White \& Basri 2003). 
We calculated global RV dispersions for all brown dwarfs and stars in Cha\,I 
(1.24\,km\,s$^{-1}$) and Taurus (2.0\,km\,s$^{-1}$) and found that the value for Taurus is significantly higher
than the one for Cha\,I by about a factor of two. Given that the stellar density of 
Taurus is much smaller than of Cha\,I (Oasa et al. 1999),
we conclude that 
a fundamental increase of velocity dispersion with stellar density of the star-forming region
as suggested by Bate \& Bonnell (2005) is not established observationally.

We compared the kinematic study in Cha\,I with theoretical hydrodynamical or N-body calculations of the 
embryo-ejection scenario for the formation of brown dwarfs.
That there is no significant difference between the RV dispersion of brown dwarfs and T~Tauri stars
in Cha\,I and that the differences found in the cumulative RV distributions
for both groups might be explained by systematic errors 
are both consistent with the finding of no mass dependence of the velocities in models 
by Bate et al. (2003), Bate \& Bonnell (2005) and with the finding of only small mass dependence in the 
model by Delgado-Donate et al. (2004).
However,
the observed RV dispersions in Cha\,I for both the brown dwarfs and the stars 
are much smaller than predicted by these models.
There is a difference of about a factor of ten to one hundred in stellar density between Cha\,I
and the simulated star-forming regions.
While an extrapolation of the predictions of Bate \& Bonnell (2005) 
to these smaller densities might be consistent with the empirical value measured by us for 
Cha\,I, we show that such an extrapolation is not advised, amongst others, because 
it does not, at the same time, yield consistent results for the less dense Taurus region. 

Sterzik \& Durisen (2003) and Umbreit et al. (2005) provide cumulative distributions of their 
results, which we can compare directly to our observed cumulative RV distributions.
The high-velocity tail seen by Sterzik \& Durisen (2003) is even more pronounced in the
models by Umbreit et al. (2005), who consider ongoing accretion during the
dynamical encounters. However, it is not seen in the observed RV distribution of brown dwarfs in
Cha\,I. We suggest that the brown dwarfs in Cha\,I show no high-velocity tail, 
but the other possibility is that 
we have missed the 40--50\% fast-moving
brown dwarfs in our relatively small sample comprised of nine objects.

We found that a subsample of known predominantly wide 
binaries among the T~Tauri stars studied in Cha\,I has 
(i) a smaller RV dispersion (1.0\,km\,s$^{-1}$)   
and (ii) no high-velocity tail compared to the remaining `single' T~Tauri stars 
(RV dispersion 1.4\,km\,s$^{-1}$).
This observational hint of a kinematic difference between singles and binaries
is in line with theoretical predictions by Sterzik \& Durisen (2003).

The comparison of observations in Cha\,I with theoretical calculations
had to deal with the difficulty that
the current models do not predict uniform quantities to describe the velocity distribution.
Furthermore, at the moment, their predictive power is limited by simplifications 
adopted therein, e.g. the lack of gravitational potential in N-body simulations and the 
lack of feedback processes in hydrodynamical calculations, as well as 
the fact that the latter are performed for much higher densities than found in
intensively observed clouds like Cha\,I or Taurus.

The observational constraint for the velocity distribution
of a homogeneous group of closely confined very young brown
dwarfs provided by the high-resolution spectroscopic study here 
is a first \emph{empirical} upper limit for ejection velocities.
It would be valuable to extent these observations 
to those not yet observed and/or to newly detected young brown dwarfs in Cha\,I (e.g. L\'opez Mart\'\i~et al. 2004)
and in other star-forming regions,
in order to put the results on an improved statistical basis.

\begin{acknowledgements}
I am grateful to the referee, Fernando Comer\'on, for 
very helpful comments which significantly improved the paper. 
I would also like to thank Kengo Tachihara, Pavel Kroupa, 
Matthew Bate, Eduardo Delgado-Donate, and Stefan Umbreit for interesting 
discussions and Michael C. Liu for hinting at 
double entries in a table in a previous publication.
It is a pleasure to acknowledge the excellent work of the 
ESO staff at Paranal, who took all the UVES observations the present work is based on in service mode.
Furthermore, I acknowledge support by a Marie Curie Fellowship of the
European Community program `Structuring the European Research Area'
under contract number FP6-501875.
This research made use of the SIMBAD database,
operated at CDS, Strasbourg, France.
\end{acknowledgements}

\appendix

\section{Details on the T~Tauri star sample}
\label{sect:app}

The sample of T~Tauri stars was revised in the presented work
compared to Joergens \& Guenther (2001) by identifying
five double entries under different names that correspond
to the very same objects in their Table\,2
(Sz\,9\,$\equiv$\,CS~Cha; Sz\,11\,$\equiv$\,CT~Cha; Sz\,36\,$\equiv$\,WY~Cha;
\mbox{Sz\,41\,$\equiv$\,RX\,J1112.7-7637}; Sz\,42\,$\equiv$\,CV~Cha).
Furthermore, two stars (Sz15\,/\,T19 and B33\,/\,CHXR25) of the previous T~Tauri star sample
were revealed as foreground stars by Luhman (2004) and were rejected.
Moreover, RV measurements by Walter (1992)
of several T~Tauri stars in Cha\,I were not considered for the
previous kinematic study and were taken into account for the
revised version presented here.
For CHX18N, the RV measured by Walter (1992)
is significantly discrepant with the measurement of
Covino et al. (1997), thus hinting at a long-period spectroscopic binary.
For the earlier publication, only the measurement by Covino et al. (1997)
of 19.0$\pm$2.0\,km\,s$^{-1}$ for this star was taken into account, which made it an outlier in the
T~Tauri star sample, while
the paper on hand also takes the previously overlooked RV determination of 13$\pm2$\,km\,s$^{-1}$
by Walter (1992) into account.
Therefore, the new mean RV for this object results in a narrower RV dispersion of the whole sample.

The kinematic study of T~Tauri stars was also revised by
an improved analysis of the UVES spectra for B34, CHXR74, and Sz23
and, in addition to this, by taking into account new UVES-based
RVs for CHXR74 and Sz23 obtained by us in 2002 and 2004.
For Sz\,23, the change in RV from 2000 to 2004 is marginal, whereas for CHXR\,74,
the discrepancy between the mean RV for 2000 and for 2004 is more than 2\,km\,s$^{-1}$,
hinting at a spectroscopic companion 
(cf. Joergens 2003, 2005b).

\end{document}